\documentstyle{article}
\title{
\begin{flushright}
{\bf\normalsize   COLO-HEP-332\\LPTHE-Orsay-93-50}\\ \end{flushright}
\bf Damaging 2D Quantum Gravity}

\author{ {\it C.F. Baillie} \\
	 Computer Science Dept. \\ University of Colorado\\ Boulder, CO 80309,
	 USA\\ \\
	 and
         \\ \\
         {\it D.A. Johnston}\\
         LPTHE\\
	 Universite Paris Sud, Batiment 211\\
         F-91405 Orsay, France$^{1}$\\
	 \\
         Dept. of Mathematics\\
         Heriot-Watt University\\
         Edinburgh, EH14 4AS, Scotland$^{2}$ }

\begin{document} \maketitle
		      {\Large \begin{abstract}
%
We investigate numerically the behaviour of damage spreading in a Kauffman
cellular
automaton with quenched rules on a dynamical $\phi^3$ graph, which is
equivalent to coupling the model to discretized 2D gravity. The model
is interesting from the cellular automaton point of view as it lies
midway between a fully quenched automaton with fixed rules and fixed
connectivity and a (soluble) fully annealed automaton with varying rules and
varying connectivity.
In addition, we simulate the automaton on a fixed $\phi^3$ graph coming
from a 2D gravity simulation as a means
of exploring the graph geometry.
\\ \\
\\ \\ \\ \\ \\ \\ \\ \\
$1$ {\it Address Sept. 1993 - 1994} \\
$2$ {\it Permanent Address}
%
			\end{abstract} }
%
  \thispagestyle{empty}
%
%
  \newpage
%
		  \pagenumbering{arabic}

Cellular automata have found applications in many area of physics
and other sciences. One particularly intriguing idea was
put foward by Kauffman in the 1960s \cite{1},
who suggested that
a random Boolean network might serve as a model for cell
differentiation. This predated the more recent explosion
of papers on the subject that was largely inspired by the
work of Wolfram \cite{2}. In the Kauffman model each of $i = 1,2, \ldots , N$
lattice sites
contains a spin which can take the values $\pm 1$ (or equivalently a
Boolean variable that can take the values $0,1$) and the spins evolve
according to
\begin{equation}
\sigma_i (t + 1 ) = f_i \left( \sigma_{j_1}(t) , \sigma_{j_2} (t), \ldots ,
\sigma_{j_k} (t) \right)
\label{e1}
\end{equation}
where the functions $f_i$ are chosen independently and
each site $i$ has $K$ specified inputs sites
$j_1(i),  \ldots , j_k (i)$.
The original biologically motivated model of Kauffman chose the $f_i$
randomly at $t=0$ with a probability $p = 1/2$ for $f_i = +1$ and
a probability $1 - p= 1/2$ for $f_i = -1$.
for each site and also chose the $j_1(i), j_2(i), \ldots , j_k (i)$
inputs
randomly from the $N$ lattice sites at $t=0$. This
version of the model thus
displays {\it quenched} randomness as both the functions $f_i$ and the input
sites are fixed and has been called
``infinite dimensional'' because of the non-locality
of the inputs. The model must display periodic behaviour for finite $N$
as there are only $2^N$ possible spin configurations, so one can enquire
about the length and number of limit cycles as well
as quantities such as the Hamming distance between
different configurations $1$ and $2$, $D_{12} (t) $, defined as
\begin{equation}
D_{12} (t) = {1 \over N} \sum_{i=1}^N \left( \sigma^1_i (t) - \sigma^2_i (t)
\right)^2
\label{e2}
\end{equation}
where $\sigma^1$ is a spin in configuration $1$ and $\sigma^2$ is
a spin in configuration $2$.
The application to genetics arises from identifying the stable limit cycles
with
different stable genotypes.

It was found (numerically) \cite{3} that for
$K \le 2$ the length of cycles was small, of order $\beta(K) N^{1/2}$,
whereas for $K \ge 3$ it increased as $\exp ( \alpha (N) N )$, which
implied a critical $K_c$ between $2$ and $3$, with corresponding singular
behaviour in $\alpha(K)$ and $\beta(K)$. The $K \le 2$ phase was
called a frozen phase and had
\begin{equation}
\lim_{ t \rightarrow \infty} D_{12} (t) = 0
\end{equation}
for any starting configurations $1$ and $2$, whereas the $K>2$ phase was
chaotic, having
\begin{equation}
\lim_{ t \rightarrow \infty} D_{12} (t) = d_{final}
\end{equation}
with the final distance $d_{final}$ independent of the initial distance.
A simple annealed approximation to the model was solved by Derrida and
Pomeau \cite{4} and captured these features
giving a critical value of $K=2$.
One can think of implementing the approximation
in a simulation by changing both the
functions and the inputs and each time step,
which amounts to neglecting all the correlations.

In \cite{5}
Derrida and Stauffer investigated
a finite dimensional variant of the model
in which the $K$ inputs were chosen from the neighbours
of a site on various lattices. They varied the probability
$p, (1-p)$ for choosing $f_i = + 1, (-1)$, which can be shown to give
the condition
\begin{equation}
2K p_c ( 1 - p_c ) = 1
\end{equation}
for the transition between the
frozen and chaotic phases, where $p$ is now used as a control parameter
rather than $K$. Derrida and Stauffer pointed out that the
Kauffman model with varying $f_i$ on a $d$ dimensional lattice
was equivalent (as far as distances and overlaps
between configurations were concerned)
to $d+1$ dimensional directed percolation \cite{6}
on the appropriate lattices. The qualitative properties
of this annealed model were similar to the infinite
dimensional case, although the numerical values of thresholds and
exponents were different.

They also investigated numerically
a version with fixed $f_i$ and found rather different
behaviour from the infinite dimensional case, with the final distance
in the chaotic phase {\it depending} on the initial distance (cf
equ.(4)). The behaviour is summarized as:
\begin{eqnarray}
\lim_{D_{12} (0) \rightarrow 0} D_{12} (\infty) &=& Q(p) > 0, \ \ if  \ \
p > p_c
\nonumber \\
\lim_{D_{12} (0) \rightarrow 0} D_{12} (\infty) &=&  0, \ \ if \ \ p < p_c
\nonumber \\
\lim_{D_{12} (0) \rightarrow 0} { D_{12} (\infty)  \over D_{12} (0)  }
&=&  \chi (p) , \ \ if \ \ p < p_c
\end{eqnarray}
where the susceptibility $\chi(p)$ diverges at $p=p_c$. The strategy adopted
in \cite{5} was to look at the time variation of the distance between two
configurations, where one of the initial configurations
contained some ``damaged'' spins that were different to the other reference
configuration. It is possible to think of
the transition from the frozen phase to
the chaotic phase as damage percolating through the lattice, which allows
another estimate of $p_c$ if one neglects spin-spin correlations,
namely
\begin{equation}
2 p_c ( 1 - p_c) = x_c
\end{equation}
where $x_c$
is the bond percolation threshold for the lattice.
Numerical simulations of the quenched model on various lattices \cite{7,7a}
showed that frozen and chaotic phases existed on square, triangular and
various cubic lattices, whereas the honeycomb (hexagonal) lattice
possessed only a frozen phase.
It was found that estimates of $p_c$ were subject to strong finite size
effects, with only the square lattice value being pinned down with precision.
In addition, various interesting
quantities such as the fractal dimension of the cluster of damaged
spins at threshold were measured. The analogy with percolation for the
quenched model seems to be rather less solid than the annealed version,
as various thresholds and fractal dimensions differ from the percolative case
\cite{7a}.

The work of \cite{5,6,7} is much closer in spirit to standard
statistical simulations of spin models on various lattices than the
earlier ``infinite dimensional'' models. It is
now well known both from analytical work \cite{8,9}
and numerical work \cite{10} that the critical behaviour of
spin models such as the Ising model can be
modified by placing them on appropriate dynamical (connectivity)
lattices, where the spins interact with the lattice and vice-versa.
The methods of \cite{8} can also be adapted to solve bond
percolation on dynamical lattices by treating it as the limit
of a $q$ state Potts model as $q \rightarrow 1$ \cite{11},
again leading to different critical behaviour from fixed lattice
counterparts.
In the light of the different behaviour of spin models on dynamical lattices
and the relation between percolation and damage spreading in the finite range
Kauffman models
a natural question to ask is
whether the behaviour of the quenched rule automaton is also changed by placing
it
on a dynamical lattice,
which would then correspond to annealing the
inputs. Such a model lies halfway between the (soluble) model of Derrida
and Pomeau with annealed rules and inputs and the models with
quenched rules and inputs simulated by Derrida and Stauffer,
and might {\it a priori} lie in either universality class. There has been
some scepticism of the relevance of quenched rule Kauffman automata with
nearest neighbour interactions on regular lattices for biological applications
\cite{11a},
on the grounds that gene interactions are known to have considerable temporal
and spatial variability. In view of this, our model might even be considered to
be as
appropriate as the original Kauffman infinite range model since we have
included
both
spatial and {\it temporal} variation in using a dynamical lattice, whilst
preserving a
characteristic fixed rule for each lattice point or ``gene''.

A second motivation for performing a
simulation on a  dynamical lattice comes from 2D gravity: in \cite{12}
percolation on triangulations arising from simulations of 2D gravity
coupled to matter with various central charges was used to investigate
the geometry of the triangulations.
In view of the links between damage spreading in the
Kauffman model and percolation one might also hope to learn something
about discretized 2D gravity,
at least qualitatively, from simulating the Kauffman model on suitable
graphs. The geometry of the graphs is certain to be reflected in some of
the features of the damage clusters and their manner of spreading.
There is no back reaction from the automaton spins on the lattice
because we change the connectivity without reference to the spins
so we will, in effect, be investigating the lattices of pure
2D gravity with no matter.
For the case of pure 2D gravity the appropriate choice of graphs
would be
either $\phi^3$ graphs or their dual triangulations as in \cite{12}.
As there is {\it no} transition for a quenched rule Kauffman automaton
on a regular fixed connectivity $\phi^3$ lattice - the honeycomb lattice,
we choose to simulate $\phi^3$ graphs to see if the introduction of dynamical
connectivity gives a transition.

In the current work
we use the methods employed \cite{10} in our simulations of Potts models
coupled
to 2D gravity
where $\phi^3$ graphs
of spherical topology with $N=10000$ points were generated.
To avoid unnecessarily singular graphs, tadpoles and self
energy bubbles were not allowed (ie no rings of length one or two
in the graphs). The dynamical connectivity was implemented
by performing a series of local ``flip'' moves, which are
the dual of those employed on triangulations, and can be shown to be ergodic
\cite{12a}.
We found that it was a
good rule of thumb to carry out $NFLIP=N$ local ``flip'' moves on the lattice
in the spin model simulations between spin updates
in order to ensure sufficient coupling to 2D
gravity so we do the same between the automaton time steps. The rules $f_i$
are chosen at random for each graph point at the start of each run,
using the probabilities $p$ and $1-p$ for ``up'' and ``down'' rules.
As the region $p > 0.5$ is equivalent to $p \le 0.5$ on inversion of $p$ and
$1 - p$ we only simulate up to $p = 0.5$.
The $f_i$ are then held
fixed during the run, whereas the flips ensure that the inputs are shuffled
between each automaton update, which is performed according to equ.(1)
simultaneously
for each spin across the entire graph. We look at the evolution of the Hamming
distance
in equ.(2) for configurations with $4$, $100$, $400$ and $1000$ differing
initial spins
over $500$ automaton timesteps and repeat this process $50$ times for each $p$
to
obtain averaged values and errors.
This proved sufficient to obtain reasonable
error bars. The initial spin configurations
were taken as cold (ordered) starts and the initial sites
to be damaged were chosen at random on each run.
We also simulate the automaton on {\it fixed} 2D gravity graphs, where we
switch off the flips in between the automaton updates. In this case we
expect to learn something about the geometry of a given $\phi^3$ graph
by comparing the results with other fixed graphs, such as the honeycomb graph.

In Fig.1 we have plotted the time evolution of the Hamming distance for $p =
0.10$ and $p = 0.40$
on both fixed and dynamical $\phi^3$ graphs for a starting damage of $100$
to give some idea of the general behaviour. One qualitative fact is immediately
clear
from the graph: there is no transition as $p$ is varied for the fixed $\phi^3$
graph,
whereas the damage clearly spreads to reach an asymptotic value for $p=0.4$
when
the flips are
switched on. Thus the fixed 2D gravity graph, a highly disordered $\phi^3$
graph
with a complicated internal structure, behaves like its smooth $\phi^3$
honeycomb counterpart
in so far as it displays no transition to a chaotic phase at larger $p$ values.
The most interesting observation for the {\it dynamical} graph simulations,
on the other hand, is how close
they come to the analytical results
for the totally annealed automaton of Derrida and Pomeau. For a totally
annealed
Kauffman model with
$K=3$, which is also the value for the dynamical $\phi^3$ graphs here, one
expects a critical value of
\begin{equation}
p_c = 1/2 - 1/ (2 \sqrt{3}) \  \ \ \simeq 0.21
\end{equation}
and an asymptotic value of the damage $d_{final} \simeq 0.38$ at $p = 0.5$
that is independent of the starting damage. In Fig.2 we plot the asymptotic
value of the
damage after 500 time steps against $p$ for the various starting damages,
from which it is clear that the automaton is almost certainly behaving
identically to the fully annealed model. There is
a transition from a frozen to a chaotic phase in the region of $0.21 -
0.22$ and the final damage is independent of the starting damage,
for all but the smallest starting damage, which is
probably too small to avoid an atypical choice of starting sites.
We find
$d_{final} \simeq 0.366(1)$, which is a little lower than the analytical
value. The changes in connectivity in the simulation thus appear
to be sufficient to destroy the correlations between the spin values
and allow the application of the annealed approximation.

The results for the asymptotic values of the damage for a fixed
2D gravity graph are shown in Fig.3, clearly demonstrating the absence
of a transition in this case. The bond percolation threshold on 2D gravity
graphs was calculated to be $x_c \simeq 0.78$ in \cite{11}, so
the formula $2 p_c ( 1 - p_c ) = x_c$ of \cite{5} correctly predicts
no transition for the quenched Kauffman automaton in this case
(as it does for the quenched automaton on a honeycomb lattice). It is
instructive to compare the results from a similarly sized honeycomb
lattice in Fig.4 with those in Fig.3. The curves for the final damage
vs $p$ in Fig.4 for a lattice of 8192 points look, in fact, as if
there is a transition. Much larger simulations however, show
convincingly that a transition is absent
on a honeycomb lattice \cite{7}, in particular by regarding
the behaviour of the final damage as the initial damage is varied.
{}From comparing Fig.3 and Fig.4 we can therefore deduce that it
is a lot easier to see the absence of damage spreading
on the 2D gravity $\phi^3$ lattice than
on the honeycomb (ie regular $\phi^3$) lattice.
This suggests that the spreading of a damage cluster encounters
more obstacles on the 2D gravity graph than the honeycomb.

Such a conclusion is consistent with the behaviour observed for the
critical slowing down of the magnetization in cluster algorithms
for spin models on dynamical surfaces, which turned out to be
anomalously large \cite{13}. It was
suggested in \cite{13} that this was due to the inhibition of cluster
growth from bottlenecks and ``baby-universes''
in the graph geometry that only changed slowly
due to the local ``flip'' graph update. Such bottle-necks and baby universes
would also trap the damage clusters in the quenched Kauffman model
and account for both the much clearer absence of a transition than
on the honeycomb lattice and the much smaller final damage on
the fixed 2D gravity graphs compared with
the honeycomb lattice. The relative smallness of the
clusters is also an indication of the larger fractal dimension
of the 2D gravity graphs, which appears to be close or equal to 3
\cite{13a}.

In summary, we simulate a quenched rule Kauffman automaton
on dynamical $\phi^3$ graphs (as used in 2D gravity simulations) and
find that the annealed results of Derrida and Kauffman are applicable.
A simulation on a fixed $\phi^3$ graph on the other hand,
although there is no transition to a chaotic phase, gives
some indication of the complicated internal geometry that is present
in such graphs. It would be interesting to look at the crossover between
the effectively annealed behaviour on dynamical graphs and that on
fixed graphs as the number of flip updates per automaton update
was reduced, as well as examining the relation between damage spreading
and thermodynamic quantities \cite{14} for  more standard spin models
on dynamical lattices.

This work was supported in part by NATO
collaborative research grant CRG910091.
CFB is supported by DOE under
contract DE-FG02-91ER40672, by NSF Grand Challenge Applications
Group Grant ASC-9217394 and by NASA HPCC Group Grant NAG5-2218.
DAJ is supported at LPTHE
by an EEC HCM fellowship, EEC HCM network grant and an Alliance grant.
The simulations were carried out on
workstations at Heriot-Watt University and the University of
Colorado.

\bigskip
\bigskip
\centerline{\bf Figure Captions}
\begin{description}
\item[Fig. 1.] The damage vs time for $p=0.1, 0.4$
on both fixed and dynamical $\phi^3$ graphs with fixed rules.
\item[Fig. 2.] The final damage vs p for various initial
damages on a dynamical $\phi^3$ graph with fixed rules.
The various initial damages are indicated.
\item[Fig. 3.] The final damage vs p for various initial
damages on a fixed $\phi^3$ graph with fixed rules.
The various initial damages are indicated.
\item[Fig. 4] The final damage on a similarly sized
honeycomb lattice, again with fixed rules.
The various initial damages are indicated.
\end{description}
\end{document}